\documentclass[a4paper,10pt,twocolumn,preprint]{elsarticle}

\usepackage{mathrsfs,amsmath}
\usepackage[numbers]{natbib}
\biboptions{numbers,sort&compress}

\journal{Computational Science}

\begin{document}

\twocolumn[{\begin{frontmatter}

\title{Studies of accurate multi-component lattice Boltzmann models on benchmark cases required for engineering applications}

\author{Hiroshi Otomo}\ead{hotomo@exa.com}\author{Hongli Fan, Yong Li, Marco Dressler, Ilya Staroselsky, Raoyang Zhang, Hudong Chen }

\address{Exa Corporation, 55 Network Drive, Burlington, Massachusetts 01803, USA}
\cortext[mycorrespondingauthor]{Corresponding author at Exa Corporation}

\begin{abstract}
\ We present recent developments in lattice Boltzmann modeling for multi-component flows, implemented on the platform of a general purpose, arbitrary geometry solver PowerFLOW. Presented benchmark cases demonstrate the method's accuracy and robustness necessary for handling real world engineering applications at practical resolution and computational cost. The key requirements for such approach are that the relevant physical properties and flow characteristics do not strongly depend on numerics. In particular, the strength of surface tension obtained using our new approach is independent of viscosity and resolution, while the spurious currents are significantly suppressed. Using a much improved surface wetting model, undesirable numerical artifacts including thin film and artificial droplet movement on inclined wall are significantly reduced.
\end{abstract}

\begin{keyword}
\texttt{Multi-component flow \sep Lattice Boltzmann}
\MSC[2010] 00-01\sep  99-00
\end{keyword}

\end{frontmatter}
}]

\section{Introduction}
\label{sec:Intro}

Recently, there has been increased interest in engineering applications of multi-component flow simulation with the lattice Boltzmann (LB) method, because of its advantages for complex geometry and turnaround time efficiency. The lattice Boltzmann method (LBM) is based on the kinetic theory which allows to construct physical models from microscopic as well as from macroscopic viewpoints. 

While multi-component LBM's have shown promising results on a large number of academic cases, numerical accuracy and stability still represent challenges under extreme conditions such as coarse resolution and low viscosity. Actually, it is these conditions that are likely to be encountered in many engineering applications. In this paper, issues with currently existing schemes and models are pointed out, and an improved LB scheme is tested.

The paper is organized as follows. In Sec. \ref{sec:fund_formalism}, the standard LBM for multi-component flow is briefly reviewed. In Sec. \ref{sec:basic_func}, issues related to the basic functionality are specified, and an improved scheme is proposed and tested via the simulation of a two-dimensional droplet. In Sec. \ref{sec:wboundary}, cases with wall boundaries are discussed and typical issues associated with boundary models are pointed out. A new boundary model is tested on some benchmark cases. We discuss results in Sec. \ref{sec:summary}. In this paper, all physical quantities are written in lattice units, and the discrete lattice time and space increments are $\Delta x = \Delta t = 1$.

\section{Multi-component lattice Bolzmann method}
\label{sec:fund_formalism}
The commonly used LB equation for multi-component flow can be written as:
\begin{equation}
\label{eq:basic_LBM_eq}
 f_{i}^{\alpha} \left( \boldsymbol{x}+\boldsymbol{c}_{i} \Delta t, t+\Delta t
 \right) - f_{i}^{\alpha} \left( \boldsymbol{x} , t \right) 
= \mathscr{C}_{i}^{\: \alpha} + \mathscr{F}_{i}^{\alpha},
\end{equation}
where $\alpha$ stands for different components (species), $\boldsymbol{c}_{i}$ is the discrete velocity and $\mathscr{F}_{i}^{\alpha}$ is inter-component interaction force \cite{1993_Xiaowen}. The D3Q19 lattice model \cite{1992_Qian} is used so that $i$ ranges from 1 to 19, and $\mathscr{C}_{i}^{\: \alpha}$ is the particle collision operator. The simplest and commonly used one is the BGK collision operator \cite{1991_Chen,1954_Bhantnagar,1992_Chen,1992_Qian} with a single relaxation time $\tau_{\alpha}$ for the  $\alpha$-species:
\begin{equation}
\mathscr{C}_{i}^{\: \alpha} = -\frac{1}{\tau_{\alpha}} \left( f_{i}^{\alpha} -
f_{i}^{eq, \alpha} \right) .
\end{equation}
The equilibrium state $f_{i}^{eq, \alpha}$ with the third order expansion is defined as:
\begin{eqnarray}
\label{eq:feq}
f_{i}^{eq, \alpha}(\rho^{\alpha}, \boldsymbol{u}) = \rho^{\alpha} w_{i} [ 1 +
  \frac{\boldsymbol{c}_{i} \cdot \boldsymbol{u}}{T_0}  + \frac{\left(
    \boldsymbol{c}_{i} \cdot \boldsymbol{u} \right)^2}{2T_0^2} \nonumber \\ 
- \frac{\boldsymbol{u}^2}{2T_0}  
+ \frac{\left( \boldsymbol{c}_{i} \cdot \boldsymbol{u} \right)^3}{6T_0^3} -
  \frac{\boldsymbol{c}_{i} \cdot \boldsymbol{u}}{2T_0^2}\boldsymbol{u}^2 ]. 
\end{eqnarray}
Here $T_0=1/3$ is the lattice temperature, $w_i$ is the isotropic weight in D3Q19, $\rho^{\alpha}$ is the density of the component $\alpha$, and $\boldsymbol{u}$ is the mixture flow velocity:
\begin{eqnarray} 
\rho^{\alpha} = \sum_i f_i^{\alpha} \\
\rho = \sum_{\alpha} \rho^{\alpha} = \sum_{\alpha} \sum_i f_i^{\alpha} \\
\boldsymbol{u} = \frac{\sum_{\alpha} \sum_i \boldsymbol{c}_{i} \cdot f_i^{\alpha}}{\rho}.
\end{eqnarray}
There exist several models that introduce local interactions between components that are responsible for separation between the components \cite{1993_Xiaowen,1994_Xiaowen}. One of the most commonly used ones is the Shan-Chen potential force:
\begin{equation}
\label{eq:comp_interaction}
\boldsymbol{F}^{\alpha, \beta} \left( \boldsymbol{x} \right) = G^{\alpha,\beta} \rho^{\alpha}
\left( \boldsymbol{x} \right) \sum_{i} w_{i} \boldsymbol{c}_{i} \rho^{\beta}
\left(  \boldsymbol{x}+ \boldsymbol{c}_{i} \Delta t \right). 
\end{equation} 
Here, the matrix $G^{\alpha, \beta}$  defines parameters which determine the strength of interaction between components. If $G^{\alpha, \alpha} = 0$, the interaction forces only exist between different components. The equation of state for each component is that of ideal gas. If $G^{\alpha, \alpha}$ is nonzero, in addition to the interaction forces between different components, there is also a repulsive force within the $\alpha$-component. As a result, the $\alpha$-component acquires the equation of state of a non-ideal gas and phase transition within that component becomes possible. In this paper, phase transitions of single components are neglected and $G^{\alpha, \alpha} = 0$.

There are several ways to apply the forcing term $\mathscr{F}_{i}^{\alpha}$. The existing approaches have the same body force representation at the first order in resolution/time step, but different at the second and higher orders. The high order difference does have significant influence on simulation quality. In this work we use the forcing term described in \cite{2012_Qli}.

The resulting fluid velocity $\boldsymbol{u}_{F}$ is the velocity averaged over pre- and post- collision steps,
\begin{eqnarray}
\boldsymbol{u}_{F}= \boldsymbol{u}+\boldsymbol{g}\Delta t /2  \\
\boldsymbol{g} = {\sum_{\alpha} \boldsymbol{g}^{\alpha}\rho^{\alpha}} / \rho.
\end{eqnarray}
where $\boldsymbol{g}^{\alpha}$ is the acceleration of the component $\alpha$ 
derived from the intercomponent force $\boldsymbol{F}^{\alpha, \beta}$: 
$\boldsymbol{g}^{\alpha}= \sum_{\beta} \boldsymbol{F}^{\alpha, \beta} / \rho^{\alpha}$.
This quantity $\boldsymbol{u}_{F}$ is henceforth called simply \emph{velocity}. 

\section{High accuracy bulk solver}
\label{sec:basic_func}

Engineering applications usually require simulations involving various material properties and flow scenarios. Due to the jump of physical characteristics at the interface between components, accurate representation and simulation of these interfaces represents a significant difficulty. There is a consensus that numerical stability and accuracy remain two major challenges in development of multi-phase/multi-component LB flow solvers. 
To ensure numerical stability, the viscosities cannot be too small, and also the viscosity ratio between different components cannot be too large. Numerical artifacts including spurious current could often contaminate flow physics near the interface region. It becomes even more challenging when the solid-fluid interaction, i.e. surface wetting, is also considered. A new LB algorithm for the multi-component flow used in this work improves these numerical issues.

Even when the interface is static, numerical artifacts could provide a source of artificial velocity, which is called spurious velocity (cf. \cite{2012_Connington}).  The proper treatment of these phenomena is recognized as one of the key requirements for accuracy and stability of the multi-component flow modeling. In previous studies \cite{2006_Xiaowen,2007_Sbragaglia}, it is pointed out that the spurious current is associated with the insufficient isotropy of the of the numerical system caused by the discretization.

Instead of BGK, we use here a regularized filter collision operator \cite{2006_Zhang}: 
\begin{eqnarray}
\label{eq:basic_LBM_eq_imp}
 f_{i}^{\alpha} \left( \boldsymbol{x}+\boldsymbol{c}_{i} \Delta t, t+\Delta t \right)
= f_{i}^{eq, \alpha} \left( \rho_{\alpha}, \boldsymbol{u} \right) \nonumber \\
+ \left(1- \frac{1}{\tau_{\alpha}} \right) f_{i}^{neq, \alpha}
+ \mathscr{F}_{i}^{\alpha}.
\end{eqnarray}
Here $\tau_{\alpha}$ is the relaxation time of the fluid component $\alpha$ that is related to the kinematic viscosity of that component $\nu_{\alpha}$ \cite{2015_Otomo}.
$f_{i}^{neq, \alpha}$ is the regularized non-equilibrium distribution function,
\begin{equation}
\label{eq:Regualize}
f_{i}^{neq, \alpha}=\Phi^{\alpha}:\Pi^{\alpha},
\end{equation}
where $\Phi$ is a regularized filter collision operator based on Hermite polynomials \cite{2006_Zhang, 2006_Xiaowen_JFM,2008_Chen} and $\Pi^{\alpha}$ is the non-equilibrium momentum flux tensor for different components. $\mathscr{F}_{i}^{\alpha}$ is the interaction body force. The general idea and relevant algorithm details of the regularization can be found in \cite{2006_Chen, 2006_Zhang, 2006_Latt, 2006_Xiaowen_JFM,2008_Chen}. Here we would like to emphasize that this filter collision operator keeps the nonequilibrium information of moments up to the desired order, for example the 2nd   order for the momentum flux and the 3rd   order for the energy flux, and removes other higher order nonequilibrium moments in the Hermite space. Such a filtering procedure could substantially reduce unphysical noise and numerical artifacts and improve numerical stability and accuracy.

As a first test of this approach, a two-dimensional static droplet is simulated with the variable initial droplet radius, $R= \left\{ 4, 8, 16, 24, 32, 48 \right\}$, and relaxation time, $\tau^{\alpha} = \left\{ 0.525, 0.55, 1.0, 1.5, 3.0 \right\}$ for each component. The simulation domain size is five times the droplet radius and the initial density for each component is 0.22. After a steady state is reached, the droplet radius is measured by fitting the hyperbolic tangent curve to the density profile.

\begin{figure*}[htbp]
  \begin{center}
    \begin{tabular}{c}
      \begin{minipage}{0.5\hsize}
        \begin{center}
          \includegraphics[clip, width=7.5 cm]{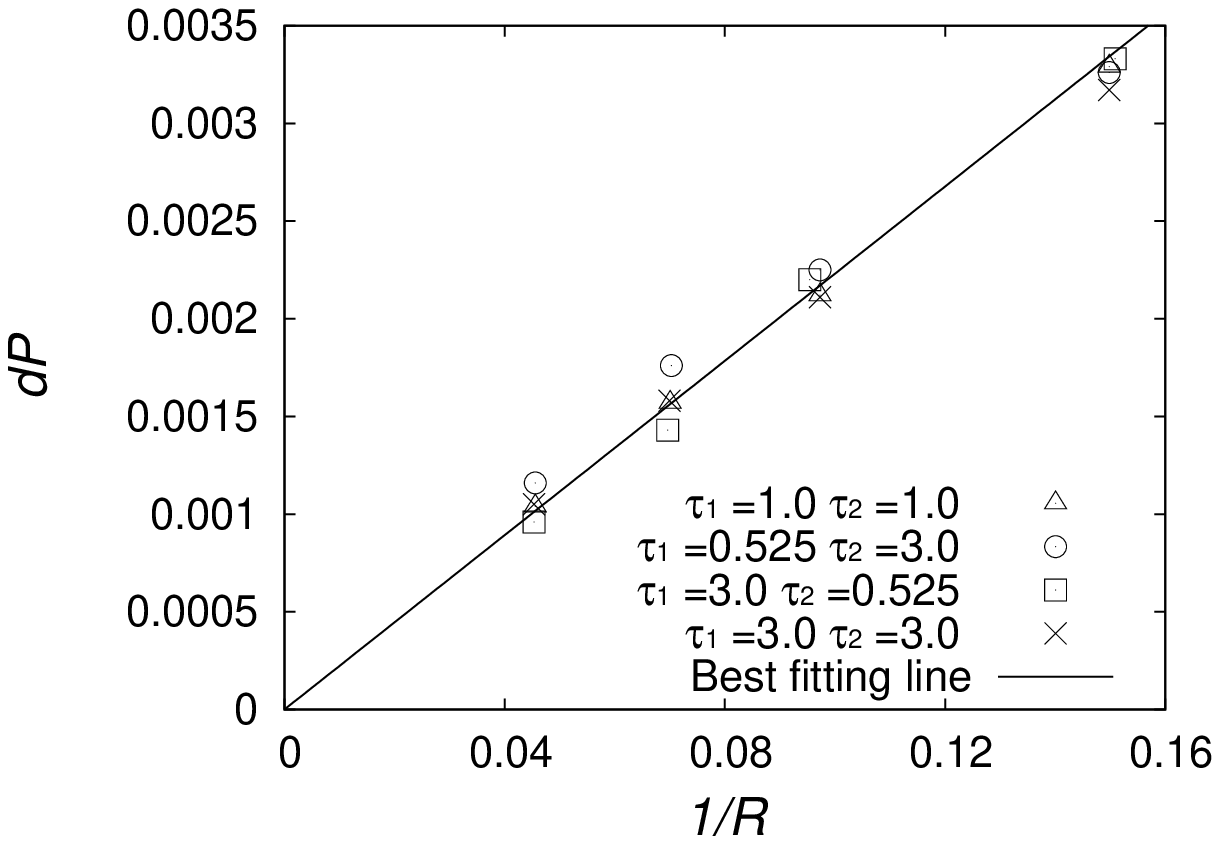}
        \end{center}
      \end{minipage}
      \begin{minipage}{0.5\hsize}
        \begin{center}
          \includegraphics[clip, width=7.5 cm]{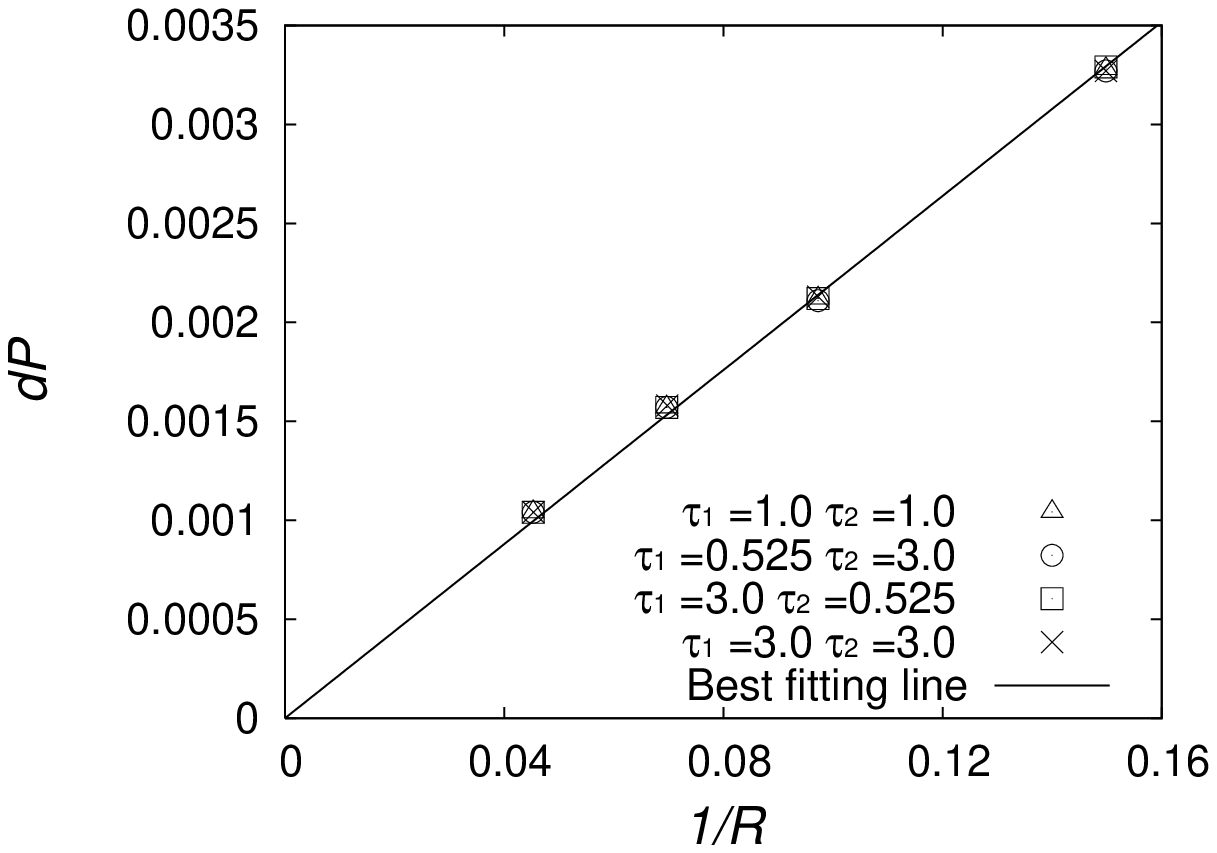}
        \end{center}
      \end{minipage}
    \end{tabular}
    \caption{Pressure difference across the droplet interface, $dP$, as a function of the inverse droplet radius, $1/R$, using the original BGK schemes (left) and new schemes (right) with four combinations of relaxation times. The subscripts 1 and 2 of denote $\tau$ inside and outside of the droplet, respectively. A line is fitted based on results with four combinations of $\tau$. }
    \label{fig:dP_vs_1R_latest_BGK}
  \end{center}
\end{figure*}

In Fig.\ref{fig:dP_vs_1R_latest_BGK}, the pressure differences across the droplet interface, $dP$, are plotted with respect to the inverse droplet radius $1/R$, using four sets of $\tau^{\alpha}$ combinations with the maximum viscosity ratio of 100. The subscripts 1 and 2 for $\tau$ denote quantities inside and outside the droplet, respectively. Results for all $\tau$ options are fitted by a line. According to the Young-Laplace law,
\begin{equation}
\label{Young-Laplace_equation}
dP=\frac{\sigma}{R} \: ,
\end{equation} 
the slope of the fitted line is the numerically achieved surface tension $\sigma$, which is independent of the viscosity and droplet size. Achieving such independence is an important first step towards simulating complex practical problems.

\begin{figure*}[htbp]
  \begin{center}
    \begin{tabular}{c}
      \begin{minipage}{0.5\hsize}
        \begin{center}
          \includegraphics[clip, width=7.5 cm]{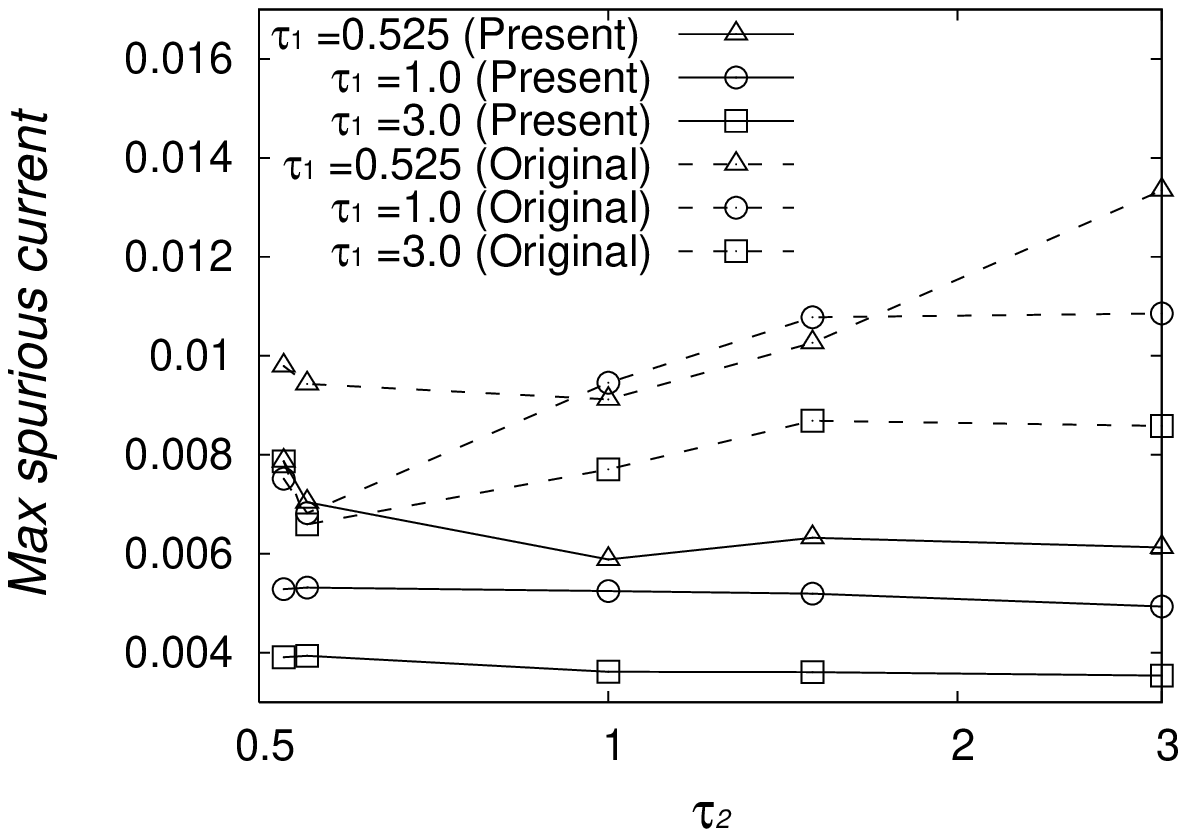}
        \end{center}
      \end{minipage}
      \begin{minipage}{0.5\hsize}
        \begin{center}
          \includegraphics[clip, width=7.5 cm]{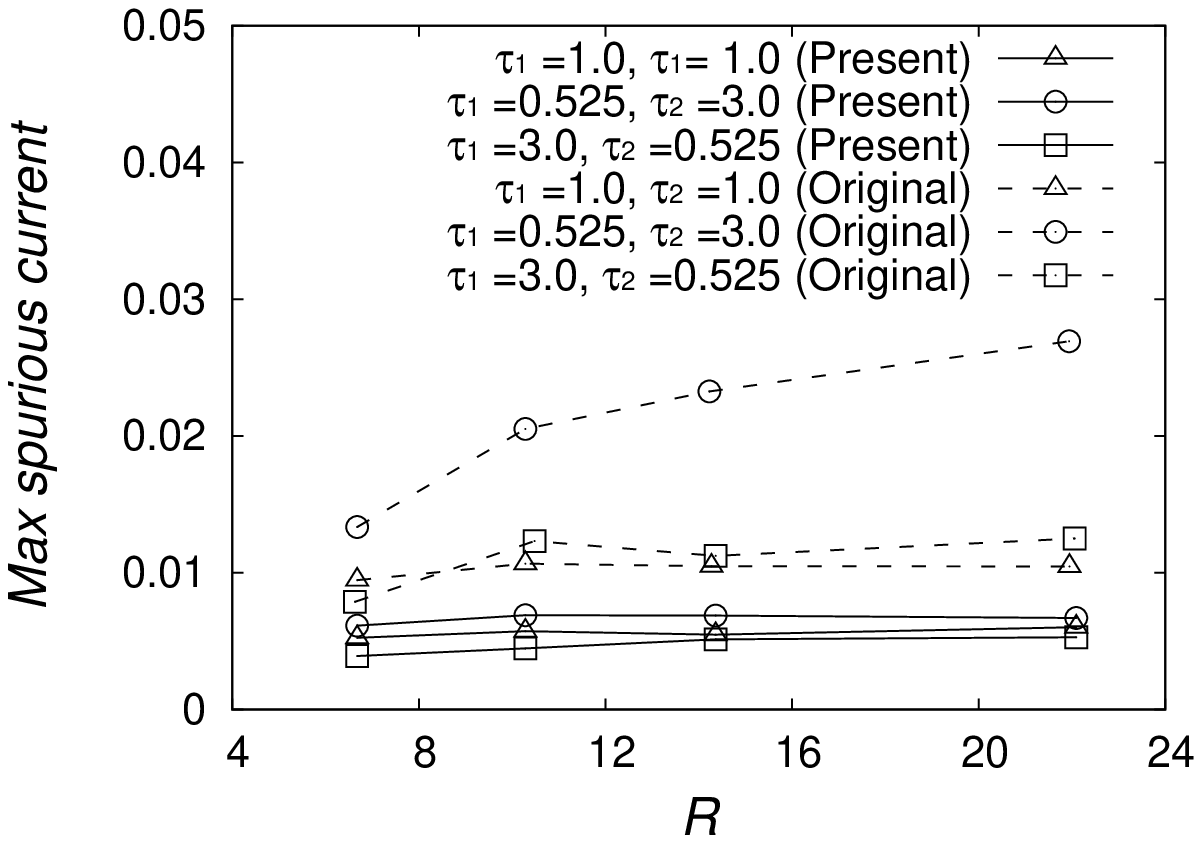}
        \end{center}
      \end{minipage}
    \end{tabular}
    \caption{Maximum spurious velocity with the original and modified scheme as a function of $\tau_2$ (left) and $R$ (right) using various viscosity combinations. In the left figure, the initial $R$ is 48. }
    \label{fig:Maxv_vs_1R_tau_latest_BGK}
  \end{center}
\end{figure*}

As mentioned above the spurious current problem is believed to be caused by insufficient isotropy of discrete schemes \cite{2006_Xiaowen,2007_Sbragaglia}. In Fig. \ref{fig:Maxv_vs_1R_tau_latest_BGK}, maximum spurious currents are plotted in terms of $\tau_2$ and $R$. In the left figure, $\tau_2$ is varied while $\tau_1$ is fixed corresponding to the initial $R = 48$. It is seen that the spurious current of the modified scheme is lower than the original one for all cases. Furthermore, with the modified version the spurious current dependence upon $\tau$ and $R$ is much reduced. As a result, one can estimate the spurious velocity quantitatively even before simulation, evaluate its effect on the main flow, and reduce numerical artifacts.

\begin{figure*}[htbp]
  \begin{center}
    \begin{tabular}{c}
      \begin{minipage}{0.5\hsize}
        \begin{center}
          \includegraphics[clip, width=7.5 cm]{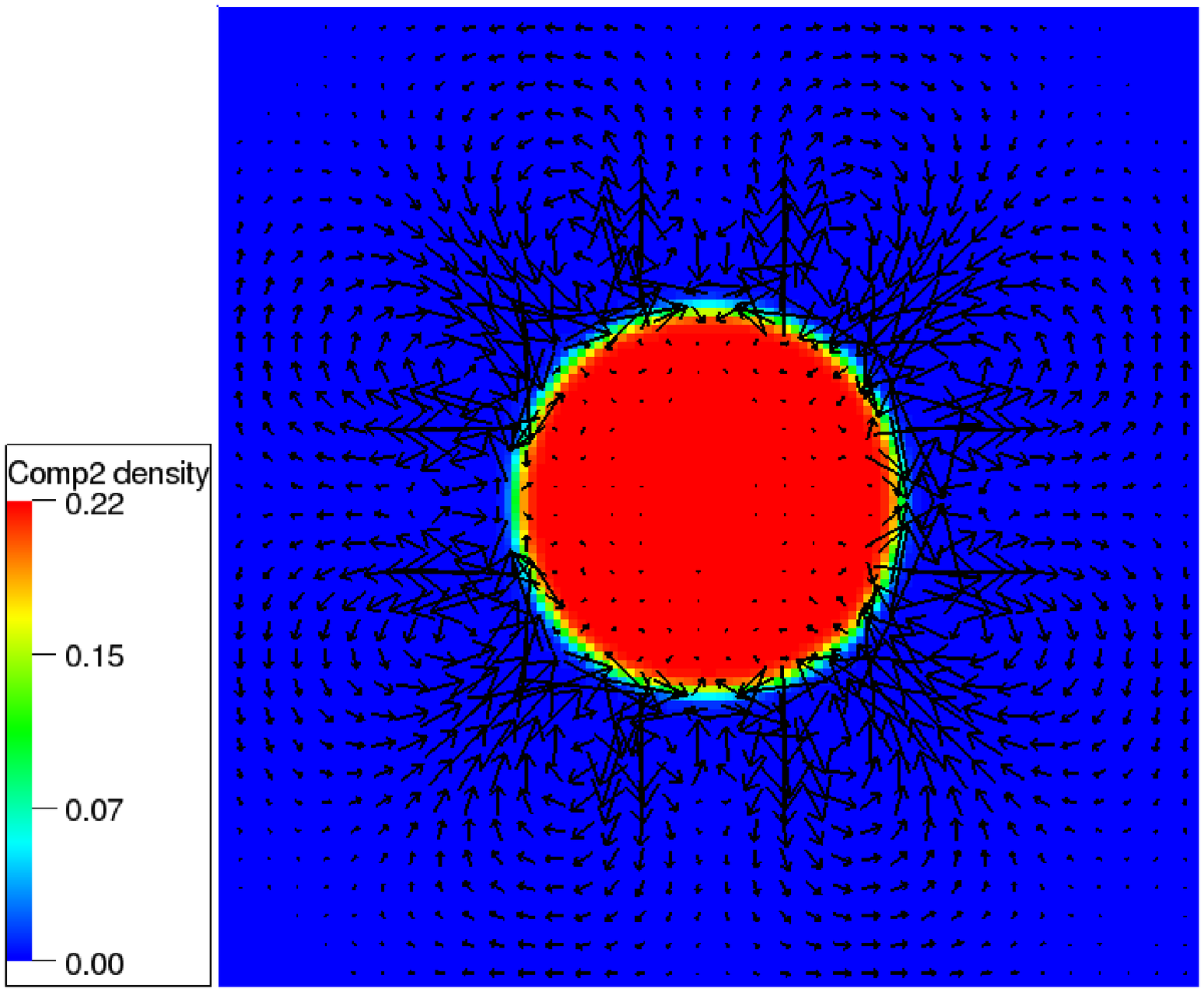}
        \end{center}
      \end{minipage}
      \begin{minipage}{0.5\hsize}
        \begin{center}
          \includegraphics[clip, width=7.5 cm]{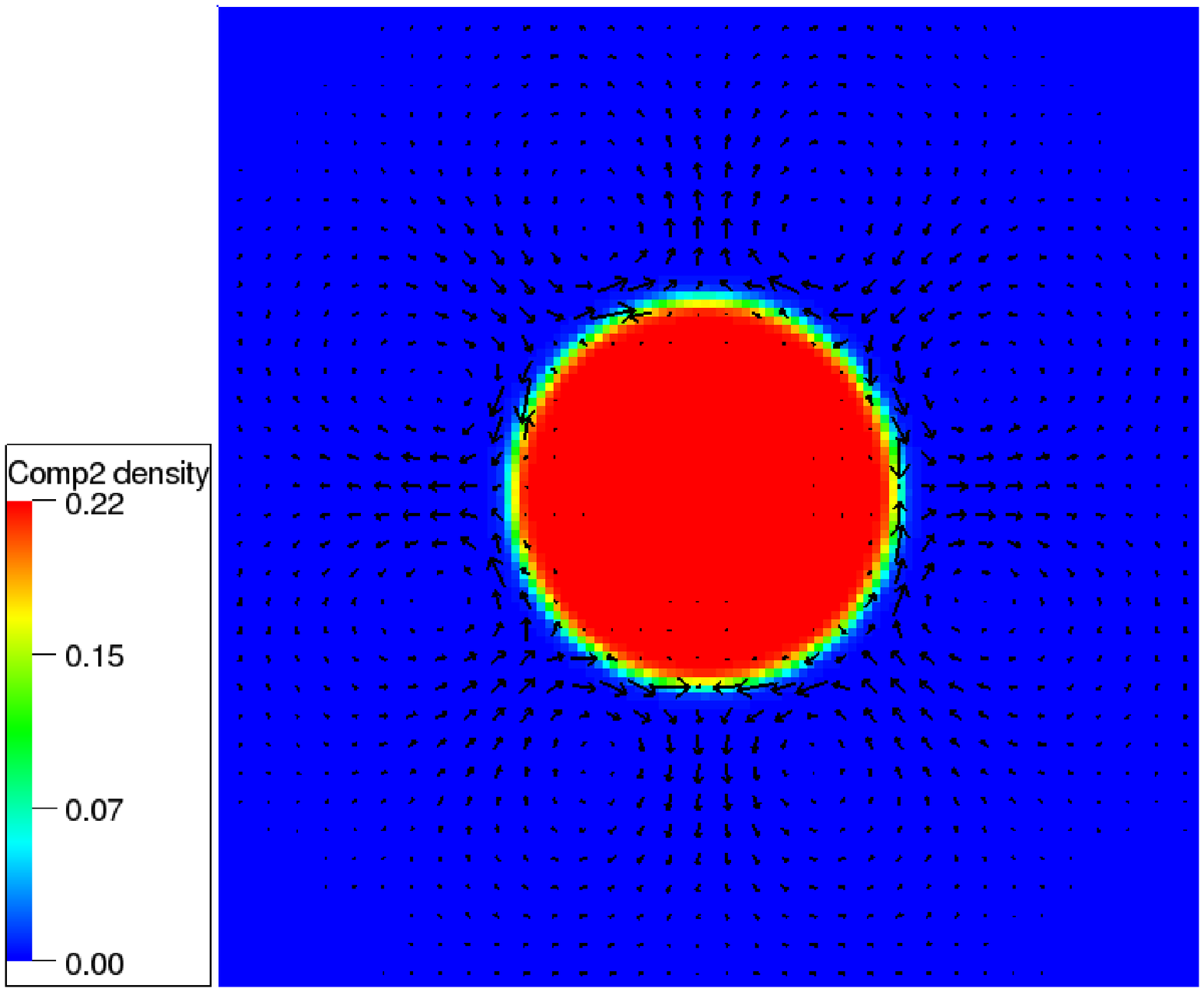}
        \end{center}
      \end{minipage}
    \end{tabular}
    \caption{Color contours of $\rho_2$ and the velocity field using the original scheme (right) and the modified scheme (left). $\tau_1 = 0.525$ and $\tau_2 = 3.0$. Initial $R$ is 48. }
    \label{fig:Maxv_vs_NuT1_0025_NuT2_25_latest_BGK}
  \end{center}
\end{figure*}

In Fig. \ref{fig:Maxv_vs_NuT1_0025_NuT2_25_latest_BGK}, the distributions of the velocity field and $\rho_2$ are presented. Here the initial $R$ is 48 and the relaxation times are $\tau_1 =0.525$ and $\tau_2 =3.0$. The results demonstrate that the new scheme  significantly reduces the spurious current while preserving the density profile and the interface thickness.

\section{Wall boundary condition}
\label{sec:wboundary}
For developed methodology of multi-component flow prediction, accurate handling of complex geometry is required. Moreover, most usage cases for multi-component  flow require accurate treatment of solid wall properties such as no-slip and wettability. 

In the standard LBM, the no-slip condition is realized with pointwise particles' bounce back model \cite{1998_Chen_Rev}. In cases with coarse resolution or low viscosity, accuracy may deteriorate. When the geometry or the local shear velocity is under-resolved, the shear stress along the wall cannot be estimated accurately. Numerical smearing could also easily contaminate flow field in the near wall region when the physical viscosity is low. As a result, the numerically simulated absolute permeability in the porous media may have significant viscosity dependence even if the Reynolds number is low enough to satisfy the Darcy's law approximately \cite{2014_Haihu}. In the previous studies \cite{2006_Pan,2015_Fattahia}, the multiple relaxation time (MRT) scheme along with the interpolated and multi-reflected boundary conditions improved this issue. However those schemes are difficult to generalize and implement for complex problems. Both of these types of boundary conditions do not conserve the local mass and require complicated local numerical interpolations in the near wall regions. Thus a more robust, accurate, and simple model is desired.

The wettability condition is often modeled by an interaction force between the fluid and the wall. A wall potential $\rho_s^{\alpha}$ is assigned to each wall to enable interactions between the solid surface and fluid particles using a point-wise concept along the lines of Eq. (\ref{eq:comp_interaction}). The correspondence between $\rho_s^{\alpha}$ and the contact angle can be  defined by simulating some test cases, after which it can be used for general cases. Although this wettability model works well for certain cases \cite{2015_Otomo}, it could sometimes generate an artificial thin film along a wetting wall. 
This film originates from slugs or droplets on the wall, and this numerical artifact is obviously undesirable because large amounts of mass may artificially diffuse and escape from the inlet or outlet boundary. Another well known undesirable issue is the artificial movement of  static droplets on inclined walls. Because of insufficient isotropy of discrete numerical schemes, even if the droplet is not subject by any driving force, the droplet may easily descend or even climb the slope. To address these issues, the improved wall boundary conditions are developed and validated.

\subsection{Modified wall models}

In our work, the bounce back model is an extension of the volumetric boundary condition proposed by Chen et al in 1998 \cite{1998_Chen,2009_Leo,2004_Yanbing,2006_Fan}, which has been extensively studied for arbitrary geometry. The main features of this model are:
\begin{enumerate}
\item Boundary surfaces are discretized into piece-wise linear surface facets in two dimensions and triangular polygons in three dimensions;
\item During the fluid dynamics calculation, the facets/polygons gather incoming particles from neighboring cells in a volumetric way;
\item At the wall, the direction of the incoming particle is flipped and the outgoing particle is constructed;
\item The outgoing particles are scattered back to neighbor cells in the similar volumetric way.
\end{enumerate} 
More details can be found in \cite{1998_Chen}, in particular a proof that conservation laws are obeyed locally as well as globally.

In addition to the correction of surface scattering described in \cite{2009_Leo}, a hybrid solid wall boundary condition for the no slip wall is proposed here in order to further reduce numerical smearing for coarse resolution simulations. The distribution function of outgoing particles $f_{i}^{out,  \alpha}$ is a combination of bounced back particles and the Maxwellian equilibrium particle distribution in accordance to viscosity values.

In our wettability model, the interaction force $\boldsymbol{F}_{w}^{\alpha,  \beta}$ is extended from the inter-component force form given by
Eq. (\ref{eq:comp_interaction}) using the volumetric boundary scheme as,
\begin{eqnarray} 
\label{eq:wall_interaction}
\boldsymbol{F}_{w}^{\alpha, \beta} \left( \boldsymbol{x} \right) = G \rho_{\alpha} \left( \boldsymbol{x} \right) \sum_{i} w_{i} \boldsymbol{c}_{i} \rho^{'}_{\beta} \left(  \boldsymbol{x}+ \boldsymbol{c}_{i} \Delta t \right) ,  
\end{eqnarray}
where $\rho^{'}_{\beta}$ is constructured in a volumetric way so that $\partial
\rho_{\beta} / \partial n = 0$ \cite{1998_Chen}. It is worth pointing out that such a volumetric wettability scheme has sufficient isotropy in complex geometry.

\subsection{Results}

The modified models are validated in order to establish that the issues specified above are improved. First, the kinematic viscosity $\nu$ dependence of the absolute permeability is investigated by simulating the simple cubic (SC) array of spheres. Second, the thin film along the wall is tested on the case of a static slug between plates. Third, the artificial movement of a droplet on an inclined wall is studied.

\subsubsection{The SC array of spheres}

The single-component gravity driven flow through the SC array of spheres is simulated. The simulation domain is set as a cube with the edge length $L=34$, containing a sphere in the center. Periodic boundaries are assigned to each pair of faced cube surfaces. The relative volume fraction $\chi=(c/c_{max})^{1/3}= 1$ and the spheres are touching each other, where $c$ is the ratio of the solid volume to the cube volume and $c_{max}$ is its maximum value, $\pi/6$ in the SC array. The choices of $\tau$ are $\left\{ 0.505, 0.51, 0.55, 0.6, 1.0, 1.5, 2.5 \right\}$. 
The gravity $g$ is in the perpendicular direction to a cube surface and $g=1.e-4$ when $\nu=0.1666$.
The value of $g$ for the other values of $\nu$ is adjusted so that $g/ \nu^2$ is kept constant. 
Such a choice keeps the Reynolds number constant for the case of the Hagen-Poiseuille flow when the channel height is fixed. Under the low Reynolds number assumption, the Darcy's law is obeyed and the flow is within the Stokes flow regime. Therefore for a given geometry, the absolute permeability $K$ should be a constant independent of physical properties as long as the flow Reynolds number is small. In our simulations, the absolute permeability is evaluated as $K = \phi \nu U /g$, where $\phi$ is the porosity and $U$ is the spatial averaged velocity.

\begin{figure*}[htbp]
  \begin{center}
    \begin{tabular}{c}
      \begin{minipage}{0.5\hsize}
        \begin{center}
          \includegraphics[clip, width=7.5 cm]{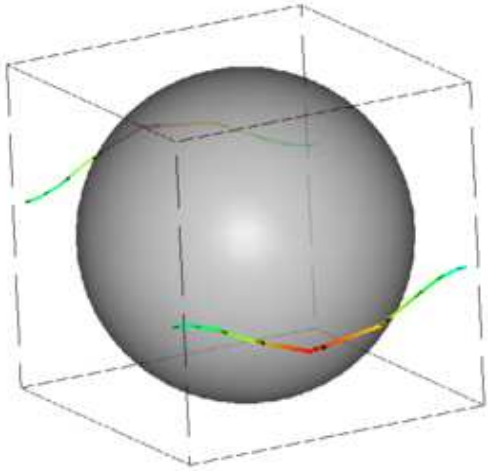}
        \end{center}
      \end{minipage}
      \begin{minipage}{0.5\hsize}
        \begin{center}
          \includegraphics[clip, width=7.5 cm]{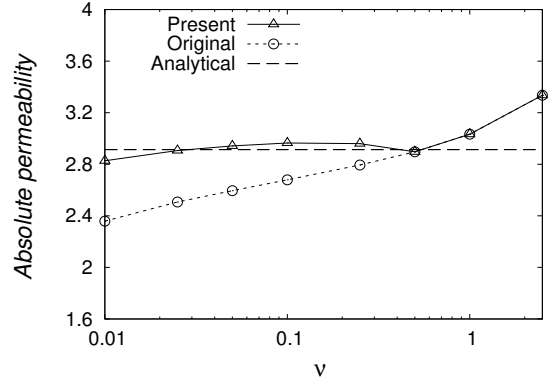}
        \end{center}
      \end{minipage}
    \end{tabular}
    \caption{Geometry and flow stream lines colored by velocity magnitude ranging from 0 to 0.0025(left), and the absolute permeability as a function of the kinematic viscosity $\nu$ with the present and original model(right). The analytical solution for permeability is shown by the dotted line.}
    \label{fig:packed_sphere_SC}
  \end{center}
\end{figure*}

A streamline in the case of $\tau=1$ is presented on the left side of Fig. \ref{fig:packed_sphere_SC}. The color represents the velocity magnitude ranging from 0 to 0.0025. Around the region where the spheres are touching, the flow path is of a narrow concave shape and therefore the flow is fast. In such a region, the resolution tends to be relatively coarse and numerical smearing could be quite pronounced when fluid viscosity is low.

On the right side of Fig. \ref{fig:packed_sphere_SC}, the simulated absolute permeability with the original bounce back and the currently presented models are shown, along  with the analytical solution obtained through the analytical dimensionless drag force from \cite{1982_Sangani,2006_Pan,2015_Fattahia}. In all simulations the Reynolds number $U L/ \nu$ is less than 0.4. The absolute permeability simulated using the original model shows a measurable dependence upon viscosity, consistent with previous studies by other researchers \cite{2014_Haihu}. On the other hand, the modified model reported in this work significantly reduces this dependence even at very low viscosities. Simulation results agree well with the analytical solution. Thus the new models can be applied for the complex geometry including sharp convex and concave shapes, while maintaining high accuracy.

\subsubsection{Static slug between flat plates}
\begin{figure*}[htbp]
  \begin{center}
    \begin{tabular}{c}
      \begin{minipage}{0.5\hsize}
        \begin{center}
          \includegraphics[clip, width=7.5 cm]{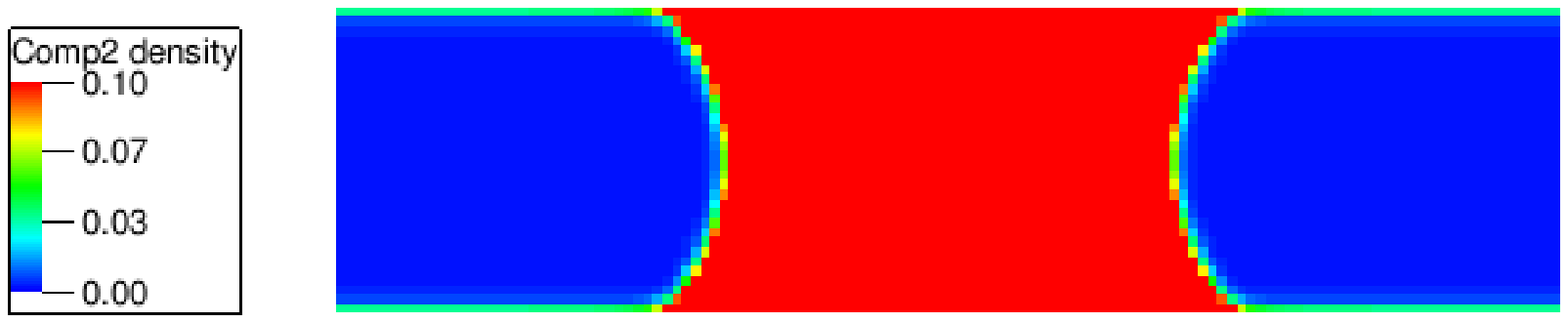}
        \end{center}
      \end{minipage}
      \begin{minipage}{0.5\hsize}
        \begin{center}
          \includegraphics[clip, width=7.5 cm]{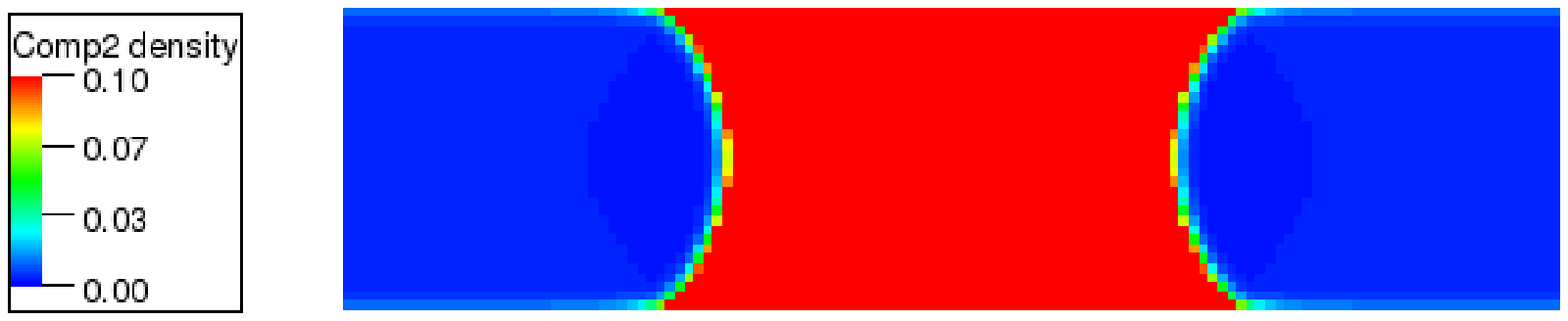}
        \end{center}
      \end{minipage}
    \end{tabular}
    \caption{Density of the second component, $\rho_2$, with the original (left) and modified wettability model (right). The channel height is 32 and $\tau$ of both components are 1.}
    \label{fig:dens_slug_between_plates}
  \end{center}
\end{figure*}

A two-dimensional static slug between flat plates is simulated under various wetting conditions. The channel height is set at 32 and $\tau$ for both components is 1. The second component is mainly located in the middle section of the channel. The wall potential $\rho^{\alpha}_{s}$ is set as $\rho^{1}_{s}=0$, $\rho^{2}_{s}=0.088$ so that the contact angle is roughly 40 degrees. In Fig. \ref{fig:dens_slug_between_plates}, the color contours show the second component density distribution with the original and modified wettability models, respectively. It is seen that the modified scheme reduces the thin 
film artifact significantly without changing the contact angle. 

\begin{figure*}[htbp]
  \begin{center}
    \begin{tabular}{c}
      \begin{minipage}{0.5\hsize}
        \begin{center}
          \includegraphics[clip, width=7.5 cm]{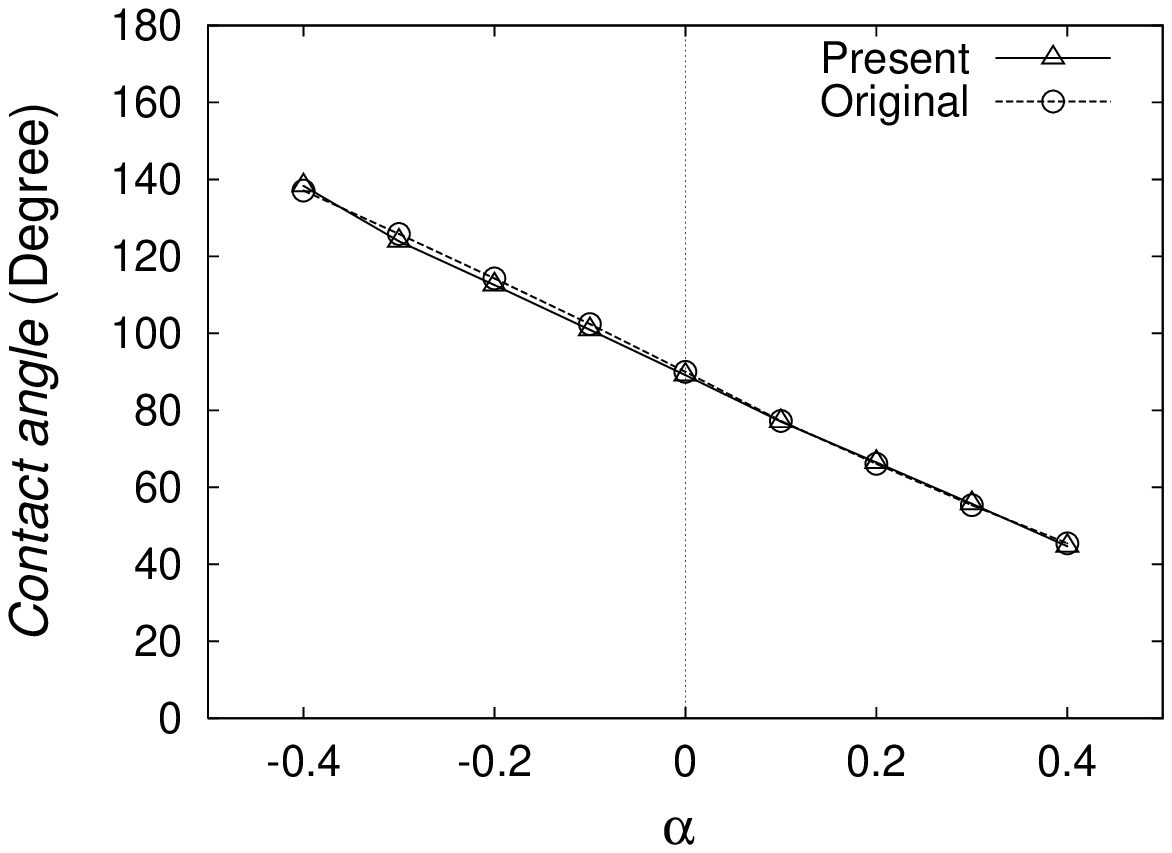}
        \end{center}
      \end{minipage}
      \begin{minipage}{0.5\hsize}
        \begin{center}
          \includegraphics[clip, width=7.5 cm]{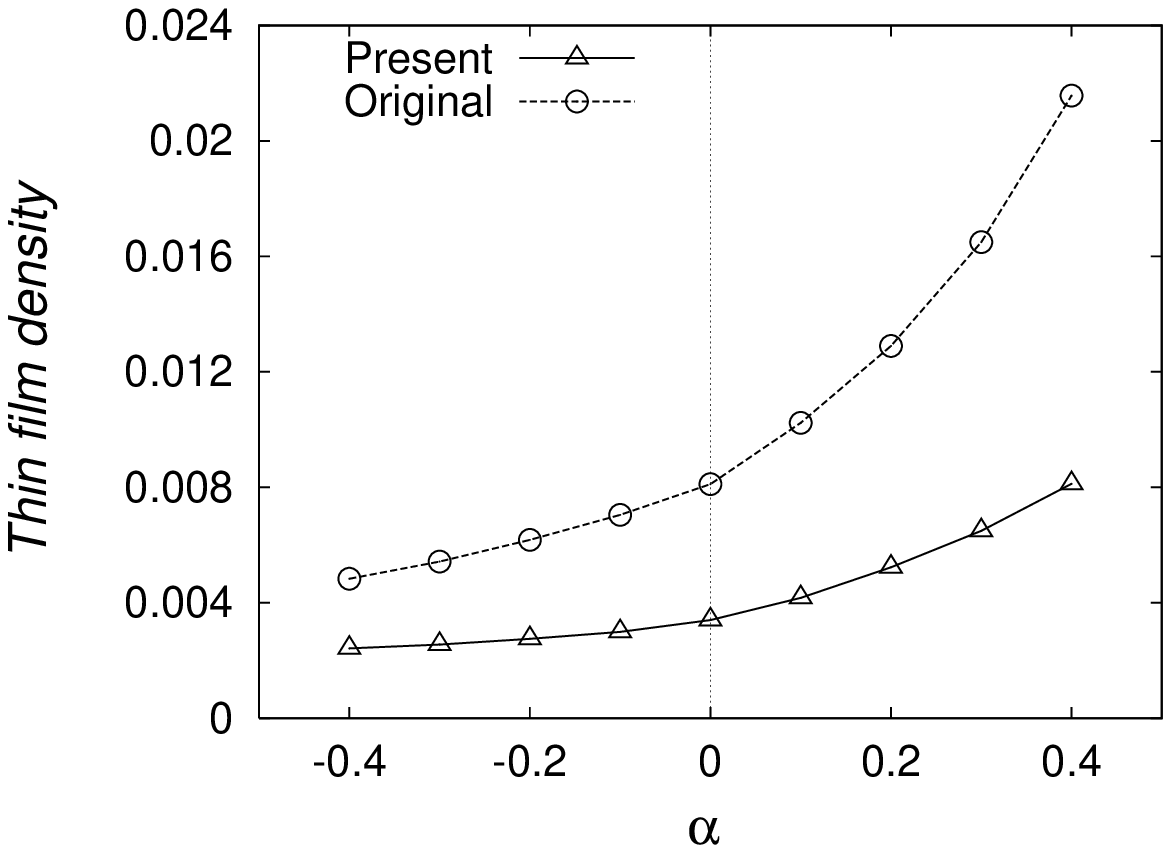}
        \end{center}
      \end{minipage}
    \end{tabular}
    \caption{The static contact angle and the thin film density as a function of $\alpha$ with original and modified models. The wall potentials $\rho_s^1$ and $\rho_s^2$ are varied following $\rho_s^1 = - \rho_0 \cdot \alpha  \cdot \Theta \left( - \alpha \right)$ and  $\rho_s^2 = \rho_0 \cdot \alpha \cdot \Theta \left( \alpha \right)$ where $\Theta$ is the Heaviside step function and $\rho_0$ is 0.22. $\tau$ of both components are 1.0.}
    \label{fig:contact_thin_film}
  \end{center}
\end{figure*}

In Fig. \ref{fig:contact_thin_film}, the contact angle and the thin film density are shown as a function of wall potential. The wall potentials $\rho_s^1$ and $\rho_s^2$ are varied as $\rho_s^1 = - \rho_0 \cdot \alpha \cdot \Theta \left( - \alpha \right)$ and  $\rho_s^2 = \rho_0 \cdot \alpha \cdot \Theta \left( \alpha \right)$. The parameter $\alpha$ is actually a dimensionless wall potential, $\Theta$ is the Heaviside step function, and $\rho_0 = 0.22$. The contact angle was measured by fitting a circle to the interface. Thin film density is detected in the first lattice cell of the simulation domain's edge. Observe from Fig. \ref{fig:contact_thin_film} that the our modification of the wall models does not change the contact angle, as intended, but at the same time reduces the thin film and wettability dependence of the thin film for all cases.

\begin{table*}[htbp]
  \begin{center}
    \caption{Droplet velocity and the ratio of droplet volume to total volume on the inclined wall with respect to component $\tau$ and the inclination angle with the original model (left) and the modified model (right). }  
    \begin{tabular}{c}
      \begin{minipage}{0.5\hsize}
        \begin{center}
            \tabcolsep = 0.14cm
	\begin{tabular}{c c c c c} \hline
	$\tau_1$ & $\tau_2$          &  Inclination    &  Droplet    & Droplet volume   \\ 
	         &                  &  angle         &   velocity & /total volume     \\  \hline
         0.55    & 1.5              &    10        &   0          &  0.077          \\ 
         0.55    & 1.5              &    30        &   6.e-4      &  0.25          \\ 
         0.55    & 1.5              &    50        &   5.e-4      &  0.21          \\ 
         0.55    & 1.5              &    70        &   3.e-4      &  0.27          \\ 
         1.0     & 1.0              &    10        &   0          &  0.078          \\ 
         1.0     & 1.0              &    30        &   4.e-4      &  0.23          \\ 
         1.0     & 1.0              &    50        &   0          &  0.20          \\ 
         1.0     & 1.0              &    70        &   3.e-4      &  0.26          \\ 
         1.5     & 0.55             &    10        &   0          &  0.057          \\ 
         1.5     & 0.55             &    30        &   3.e-3      &  0.21          \\ 
         1.5     & 0.55             &    50        &   5.e-4      &  0.19          \\
         1.5     & 0.55             &    70        &   4.e-4      &  0.24          \\ \hline
	\end{tabular}
        \end{center}
      \end{minipage}
      \begin{minipage}{0.5\hsize}
        \begin{center}
        \tabcolsep = 0.14cm
	\begin{tabular}{c c c c c} \hline	
         $\tau_1$ & $\tau_2$          &  Inclination    &  Droplet    & Droplet volume   \\ 
	         &                  &  angle       &   velocity &    /total volume      \\  \hline
	 0.55    & 1.5              &    10        &   0          &  0.25          \\ 
         0.55    & 1.5              &    30        &   0          &  0.35          \\ 
         0.55    & 1.5              &    50        &   0          &  0.33          \\ 
         0.55    & 1.5              &    70        &   0          &  0.40          \\ 
         1.0     & 1.0              &    10        &   0          &  0.24          \\ 
         1.0     & 1.0              &    30        &   0          &  0.31          \\ 
         1.0     & 1.0              &    50        &   0          &  0.28          \\ 
         1.0     & 1.0              &    70        &   0          &  0.37          \\ 
         1.5     & 0.55             &    10        &   0          &  0.24          \\ 
         1.5     & 0.55             &    30        &   0          &  0.26          \\ 
         1.5     & 0.55             &    50        &   0          &  0.26          \\
         1.5     & 0.55             &    70        &   0          &  0.34          \\ \hline
	\end{tabular}
        \end{center}
      \end{minipage}
    \end{tabular}
    \label{tab:inclined_wall}
  \end{center}
\end{table*}

\subsubsection{A droplet on the inclined wall}
\begin{figure*}[htbp]
  \begin{center}
    \begin{tabular}{c}
      \begin{minipage}{0.5\hsize}
        \begin{center}
          \includegraphics[clip, width=7.5 cm]{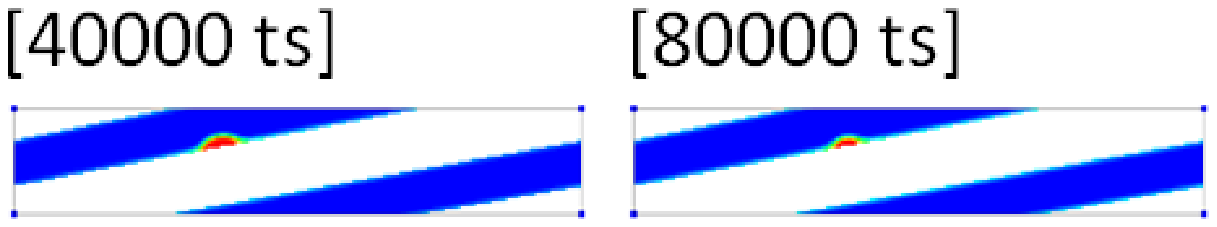}
        \end{center}
      \end{minipage}
      \begin{minipage}{0.5\hsize}
        \begin{center}
          \includegraphics[clip, width=7.5 cm]{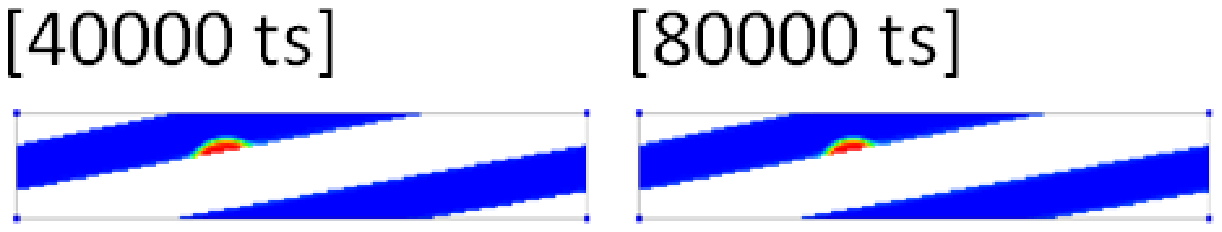}
        \end{center}
      \end{minipage}
    \end{tabular}
    \caption{The density of the second component at different timesteps with the original (left) and modified (right) models. The density range is from 0 to 0.22. The angle of the inclined wall is 10 degrees;  $\tau_1=0.55$ and $\tau_2=1.5$. No explicit driving force is applied. The periodic boundaries are enforced between the top/left and bottom/right edges.}
    \label{fig:inclined_ang10}
  \end{center}
\end{figure*}

\begin{figure*}[htbp]
  \begin{center}
    \begin{tabular}{c}
      \begin{minipage}{0.5\hsize}
        \begin{center}
          \includegraphics[clip, width=7.5 cm]{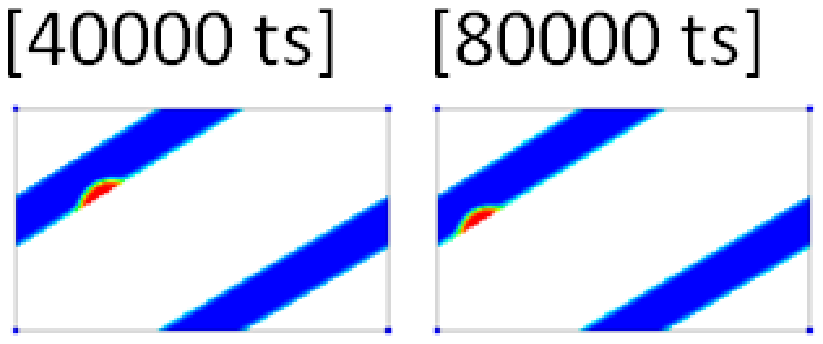}
        \end{center}
      \end{minipage}
      \begin{minipage}{0.5\hsize}
        \begin{center}
          \includegraphics[clip, width=7.5 cm]{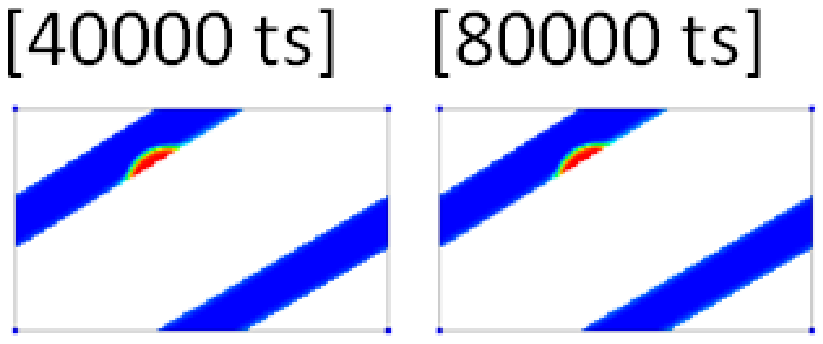}
        \end{center}
      \end{minipage}
    \end{tabular}
    \caption{The density of the second component at different timesteps with the original (left) and modified (right) models. The density range is from 0 to 0.22. The angle of the inclined wall is 30 degrees;  $\tau_1=0.55$ and $\tau_2=1.5$. No driving force is applied. The periodic boundaries are enforced between the top/left and bottom/right edges.}
    \label{fig:inclined_ang30}
  \end{center}
\end{figure*}

A two-dimensional droplet that is not subject to any explicit driving force on an inclined wall is simulated. The droplet is composed by the second component, surrounded by the first component. The channel height is 16 and the wall potential is set similarly to the slug case above. The periodic boundaries are enforced on the top/left and bottom/right edges. The inclination angles are $\left\{ 10, 30, 50, 70 \right\}$ degrees. In the cases with the same inclination angle, exactly the same initial droplet mass is set. In Fig. \ref{fig:inclined_ang10} and Fig. \ref{fig:inclined_ang30}, the results obtained with the original and modified models are compared using the inclination angles of 10 and 30 degrees, for $\tau_1=0.55$ and $\tau_2=1.5$. Fig. \ref{fig:inclined_ang10} shows that the droplet is less diffusive with the modified models, which is likely due to the reduction of the thin film effect that has been shown for the previously discussed cases. In Fig. \ref{fig:inclined_ang30} the droplet simulated with the modified model is static, as desired, while the one with the original model is moving. In Table \ref{tab:inclined_wall}, results for various $\tau$ combinations and inclination angles are presented. The droplet velocity is calculated by measuring the position change of the droplet center mass between 8.e+4 and 8.5e+4 time steps. Therefore the velocity less than 2.e-4 cannot be measured accurately and is set to 0. Utilizing the ratio of the droplet volume to total volume, the mass diffusion can be evaluated and the quantitative comparison between the two models is presented. The droplet volume is detected by adding volume where the second component density is more than 0.18. The results show that the modified model improves the artificial droplet movement and mass diffusion on the inclined wall for all tested inclinations.

\section{Summary}
\label{sec:summary}
An enhanced multi-component LB flow solver is presented whose accuracy and stability are demonstrated on a set of difficult benchmark cases. The algorithm and numerical scheme are generalized for practical applications. The surface tension achieved in the simulation is independent of fluid viscosity and resolution. The spurious current is significantly reduced. The new surface wetting scheme for complex geometry improves the near wall algorithm isotropy and the overall quality of numerical solution in the near wall region, in particular it reduces the unphysical surface thin film and mass diffusion. The model enables simulation of multi-component flows in an extended viscosity range and in complex geometry with much improved accuracy, stability, and robustness.

\end{document}